\documentstyle[twoside,fleqn,espcrc2,epsfig]{article}

\def\nostrocostrutto#1\over#2{\mathrel{\mathop{\kern 0pt \rlap
  {\raise.2ex\hbox{$#1$}}}
  \lower.9ex\hbox{\kern-.190em $#2$}}}
\def\gsim{\nostrocostrutto > \over \sim}   

\newcommand{\eref}[1]{(\ref{#1})}      

\newcommand{\N}{{\mathcal N}}
\newcommand{\Nq}{{\mathcal N}_q}
\newcommand{\Ng}{{\mathcal N}_g}
\newcommand{\Ord}{{\mathcal O}}

\newcommand{\MeV}{{\rm MeV}}
\newcommand{\GeV}{{\rm GeV}}

\newcommand{\epem}{\mbox{$e^+e^-$}}

\newcommand{\AmS}{{\protect\the\textfont2
  A\kern-.1667em\lower.5ex\hbox{M}\kern-.125emS}}
\title{Hadron multiplicity as the limit of 
jet multiplicity at high resolution\thanks{MPI-PhT/97-55, September 1997,
presented at the High Energy Physics International Euroconference
 on Quantum
ChromoDynamics, Montpellier, July 1997.} }

\author{Sergio Lupia and Wolfgang Ochs\address{Max-Planck-Institut f\"ur
Physik (Werner-Heisenberg-Institut),\\
 F\"ohringer Ring 6, D-80805 M\"unchen, Germany}%
}       
\begin{document}

\thispagestyle{empty}

\begin{abstract}
Recently exact numerical results from the evolution equation for
parton multiplicities in QCD jets have been obtained. A comparison 
with various approximate results is presented.
A good description is obtained
not only of the jet multiplicities measured at LEP-1 
but also of the  hadron
multiplicities for $cms$ energies
above 1.6 GeV in \epem annihilation.
The solution suggests that a final state 
hadron can be represented by a jet in the limit of
 small (nonperturbative) $k_\perp$ cut-off $Q_0$. 
In this description using as adjustable parameters 
only the QCD scale $\Lambda$ and  the cut-off $Q_0$, 
the coupling $\alpha_s$ can be seen to rise
towards large values above unity at low energies. 
\end{abstract}

\maketitle

\section{INTRODUCTION}

The phenomena of multiparticle production in hard processes 
require for their description within QCD both perturbative and non-perturbative
elements. Whereas the perturbative part -- at least in principle -- is
rather well understood, there are different approaches to deal with the
hadronization at large distances and also the characteristic scale for the
transition between both regimes is not uniquely defined. 
Recently, we
studied this
transition region more quantitatively in case of a simple observable,
the multiplicity of jets in an event \cite{lo2}.

At a large resolution scale $Q_c$ only a few jets are resolved  
and their rate can be calculated in fixed order perturbation theory; 
with increasing resolution more and more jets are resolved 
and in this case results can be obtained by a resummation of the
leading and next-to-leading logarithmic terms of the
perturbation theory \cite{cdotw}. In these calculations the multiplicity is
obtained in absolute normalization and the only free parameter is the QCD
scale $\Lambda$.
Increasing the resolution further  
the jet multiplicity coincides finally with the hadron multiplicity.
The jet multiplicity is defined for a given resolution cut-off parameter $Q_c$
using the well known Durham/$k_\perp$ algorithm which counts some clusters of
particles as separate if their relative transverse momenta are larger than 
 $Q_c$. In \epem one also introduces the normalized scale $y_c=(Q_c/Q)^2$ where
$Q$ denotes the total $cms$ energy. 

A  description of hadron spectra and hadron multiplicity 
has been previously obtained from the 
modified leading log approximation (MLLA) of QCD assuming a close
similarity of parton and hadron spectra with a small transverse momentum
cut-off $Q_0$ for the parton cascade (Local Parton Hadron Duality -- LPHD 
\cite{dkmt}). 
In this description the parameter $Q_0$ represents the 
confinement scale which terminates the perturbative evolution.
The multiplicity of hadrons is then found
proportional (roughly twice in the fits) to the multiplicity of partons,
also the energy dependences are  slightly different (for a recent
analysis, see \cite{lo}). One might therefore expect that with decreasing 
$k_\perp$ cut-off $Q_c$ the perturbative regime ends before  
the hadronic regime is reached at $Q_c\to Q_0$.

In our analysis we  search for  the existence of such an
additional scale. Furthermore we investigate whether we can
see from the experimental data some characteristic
features of the perturbative analysis, in particular, the strong 
variation  of the coupling $\alpha_s(k_\perp)$ when approaching the QCD scale
$\Lambda$.

\section{PERTURBATIVE ANALYSIS}

\begin{figure*}[t]
\begin{center}
\mbox{\epsfig{file=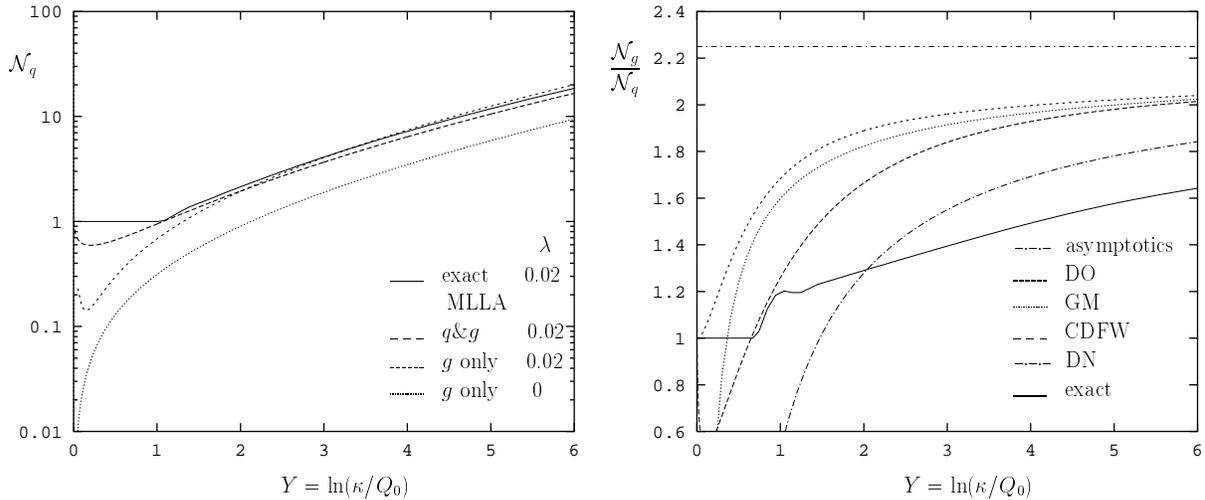,bbllx=3.5cm,bblly=13.cm,bburx=20.5cm,bbury=19.cm
,width=16.cm}}
          \end{center}
\caption{Comparison of different approximations to the master equation for
jet evolution \protect\cite{dkmt} 
as a function of the jet energy variable $Y=\ln(\kappa/Q_0)$
($\kappa=2E\sin(\Theta/2),\; \Theta=\pi/2$) : 
(a) Multiplicity in quark jets; our exact numerical solution 
\protect\cite{lo2}, 
the MLLA results from the coupled eqn. for $q\&g$ jets \protect\cite{cdotw}, 
the solution for $g$ jets only, multiplied by 4/9 and the limiting spectrum
with $\lambda=0$, multiplied by 4/9 \protect\cite{dkmt}; 
(b) Multiplicity ratio for exact
solution \protect\cite{lo2} 
compared with asymptotic value 9/4 \protect\cite{bg}, the
MLLA results using the normalization at threshold
(DO \protect\cite{do} and CDFW \protect\cite{cdotw}) and asymptotic
expansions 
(GM \protect\cite{gm} and DN \protect\cite{dn}) for $\lambda=0.02$.} 
\end{figure*}

The inclusive and exclusive characteristics of the multiparticle final
states
can be conveniently treated using the generating functional. It fulfils a
coupled system of
evolution equations (\lq\lq master equation'') in the virtuality
$\kappa=2E\sin(\Theta/2)$ with
$\Theta$ the opening angle and $E$ the energy
of the quark or gluon jet \cite{dkmt}.
It takes into account the angular ordering inside the parton cascade
as well as the running of the coupling with the transverse momentum of the 
emitted parton. For initialization of the jet evolution one takes the
condition that at the threshold of the process there is only one parton in
the jet which carries the full momentum. After appropriate functional
differentiation one obtains the coupled integro-differential equation for
quark and gluon jet multiplicities $\Nq$ and $\Ng$ \cite{lo2}.

The master equation yields the complete MLLA asymptotics, i.e. the correct
leading and next-to-leading terms in an expansion of $\sqrt{\alpha_s}$.
Furthermore it generates a perturbative expansion which allows for the
fulfilment of the initial conditions at threshold. The master equation can
be simplified by making high energy approximations already in the integral
kernel (\lq\lq MLLA evolution equation'' \cite{do}). For this simplified
equation analytic expressions can be obtained for the multiplicities in
quark and gluon jets \cite{cdotw}. Because of the high energy approximations
the behaviour near threshold has some unphysical features (multiplicities
smaller than one). To avoid this difficulty we have therefore 
solved the master equation numerically for given parameters $\Lambda$ and
 $Q_c$ (or $Q_0$).

In Fig. 1 we compare the results for the multiplicities in quark and gluon
jets for different approximations to the master equation choosing a small
value $\lambda=0.02$ for
the parameter $\lambda=\ln(Q_0/\Lambda)$ as obtained below 
from a fit to the hadron multiplicities. One can see in (a) that the 
quark jet multiplicities for the exact solution 
and the MLLA at the same $\lambda$
behave very similar already above the inelastic threshold;  the result for
the \lq\lq limiting spectrum'' at $\lambda=0$ \cite{dkmt} 
which has the limit
$\Nq\to 0$ at threshold $Y\to 0$ is smaller in absolute normalization by 
roughly a factor 2. It should be noted however that the close agreement 
within 20\% of the                                
upper three curves is special for $\lambda=0.02$ and 
disappears with increasing $\lambda$. In Fig. 1b we compare
our exact solution of the master equation
for the ratio of multiplicities in quark and gluon jets, again at
$\lambda=0.02$, 
with several approximations using and not using the
initial conditions at threshold. One can see that at LEP energies
($Y\approx 5$) the exact solution is smaller than all quoted approximations. 
 
A qualitative difference between the exact solution of the master equation
and the analytic MLLA results is 
found in the behaviour for $\lambda\to 0$: whereas
the multiplicity approaches a finite value in the latter case, it diverges 
for the exact solution  as it does in case of the double log
approximation (DLA) with \cite{dkmt}
\begin{equation}
\N\sim
K_0(A\sqrt{\lambda_c})I_1(A\sqrt{\ln(\kappa/\Lambda)})+\ldots
\label{ndla}  
\end{equation}
which at high energies and small $\lambda_c$ behaves like
$\ln \lambda_c \exp (A \sqrt{\ln (\kappa/\Lambda)})$ and exhibits a
logarithmic singularity at $\lambda_c=0$.

The behaviour near threshold can be improved by taking into account the full
perturbative result at fixed order (see, for example, ref. \cite{cdotw}).
Such results depend on the particular process. For \epem annihilation the 
$\Ord(\alpha_s)$ corrected solution has been calculated 
\cite{lo2} as
\begin{equation}
\N_{corr}^{e^+e^-}(y_c) =2 \N_q(y_c)-2 \N_q^{(1)}(y_c) +\N^{3-jet}(y_c)
\label{nepem}
\end{equation}
where $\N^{3-jet}(y_c)$ is the numerically integrated 3 jet cross section
(for running $\alpha_s(k_\perp)$) and $2 \N_q^{(1)}(y_c)$ is the corresponding
quantity in $\Ord(\alpha_s)$ from the master equation.

\begin{figure}[htb]
\begin{center} 
\mbox{\epsfig{file=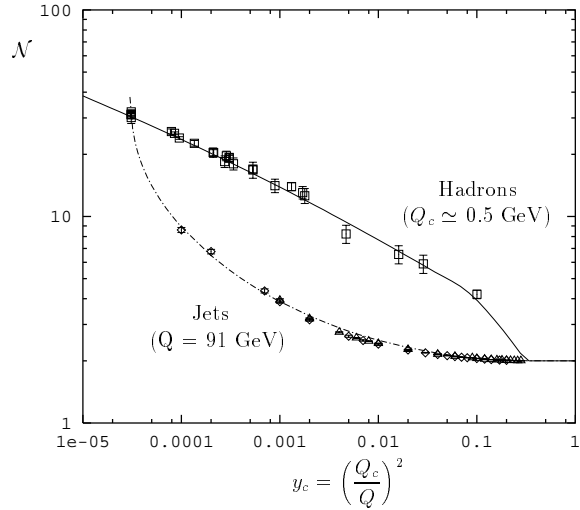,width=8.9cm,bbllx=4.cm,bblly=12.cm,bburx=21.cm,bbury=22.cm}}
\end{center}
\caption{Experimental data of the average jet multiplicities at LEP-1 
as a function of the resolution parameter $y_c$; also shown are
the average hadron multiplicities (assuming $\N = \frac{3}{2} \N_{ch}$)
at different $cms$ energies
using $Q_c$ = 0.507 GeV.
The solid (dashed) line shows the prediction in absolute normalization
 for the hadron (jet)
multiplicities obtained by 
using eq.~\protect\eref{nepem} with parameters from
eq.~\protect\eref{results}
 (figure from ref. \protect\cite{lo2}).} 
\label{figmult}
\end{figure}

\section{COMPARISON WITH $e^+e^-$ DATA}
In Fig. 2 we show the data of the average jet multiplicity at $Q$ = 91 GeV
as a function of the resolution parameter  
$y_c$ obtained with the $k_\perp$ algorithm. 
The theoretical predictions from (\ref{nepem}) for the jet data are given in
absolute normalization in terms of the single parameter $\Lambda$.
Also shown are
the data on hadron multiplicities in the energy range $Q$ = 1.6$\ldots$
91 GeV
taken as $\N_{all}=\frac{3}{2} \N_{ch}$
as a function of the
same scale parameter, now calculated as  $y_c=Q^2_0/Q^2$,  where 
the parameter  $Q_0$ corresponds to the parton $k_\perp$  
cutoff characterizing a hadronic scale according to the LPHD picture and is
obtained from a fit to the data. Another adjustable parameter here is the
overall normalization $K_{all}$ which relates the parton and hadron
multiplicities in $\N_{all}=K_{all} \N_{corr}^{e^+e^-}$.
A good description of the data
is obtained with parameters
\begin{equation}
K_{all}=1, \; \Lambda=500\pm 50~ \MeV,
\;  \lambda=
0.015\pm 0.005\,   \label{results}
\end{equation}
($\lambda=\ln (Q_0/\Lambda)$)
which correspond to the curves in Fig. 2. The fits 
describe the data within 5\% or within the errors.

A remarkable result of our analysis with improved accuracy 
is the common normalization
 $K_{all}=1$ which is possible without difficulty;
the normalization parameter $K_{all}$ is correlated with $Q_0$
and can be varied within about 30\%.
This differs from the earlier results based on the
limiting spectrum (approximate solution with $\lambda=0$) which lead to the
larger value  $K_{all}\approx 2$ (see, for example, refs. \cite{dkmt,lo}). 
This difference can be traced back
to the different normalization at threshold, see Fig. 1a. 
The solution with $K_{all}$ = 1 is natural as
it provides  the correct boundary conditions $\N=2$
at threshold for both hadrons and jets. Then
a continuous connection  between jet and hadron multiplicities
results: both are described by
the same equation (\ref{nepem}) and one is obtained from the other 
at fixed $y_c$ by changing the absolute scale $Q_c$ down to $Q_0$.
There is no evidence for another nonperturbative scale above $Q_0$.
In this description
a single hadron corresponds to a single parton of low virtuality
$Q_0$. 

An important role in this interpretation of the data is played by
 the running of $\alpha_s$.
First we note that 
in a model with fixed $\alpha_s$ the multiplicity would depend 
only on the ratio of the available scales through the variable $y_c$
but not on the absolute scales $Q$ or $Q_c$ separately; 
such a model would predict at high energy a power like
dependence on $Q/Q_c$, i.e. a straight line in Fig. 2 
for both the hadron and jet multiplicities between the two curves
shown. 

On the other hand, in case of running  $\alpha_s$ the additional
scale $k_\perp/\Lambda$ appears which, for a given $y_c$, 
leads to an enhancement of
multiplicities at small absolute scales $Q$ or $Q_c$ and hence to the
different curves in Fig. 2. The effect of the running $\alpha_s$ is most
pronounced at large  $y_c$. Here the quantity $\N-2$ is of $\Ord (\alpha_s)$
and therefore reflects the typical coupling. From the 
separation of the jet and hadron data in Fig.
2 (see also data on $\N-2$ in \cite{lo2}) one finds that around $y_c\sim
0.1$ the typical couplings should differ 
by more than an order of magnitude. 
As the coupling to produce jets at LEP energies is around $\alpha_s\sim 0.1$ 
 the coupling to produce particles at low energies is therefore
\begin{equation}
\alpha_s>1\quad {\rm at}\quad Q\sim 1.6\;\GeV.
\end{equation}
The fact that both data sets are rather well described over the full 
kinematic region where data are available 
confirms that $k_\perp$ is a good choice for the scale of $\alpha_s$  
  and also supports the LPHD concept to describe hadrons by the $k_\perp$
 cut-off $Q_0$. 
There is a certain ambiguity in the definition of $k_\perp$ which
leads to different values for $\Lambda$. The value quoted above in 
(\ref{results}) refers to the Durham-$k_\perp$ definition whereas the usual
$k_\perp$ would yield a smaller value around 300 MeV \cite{lo2}. 

Another characteristic sign of the running coupling is the strong
rise of the jet rate towards small $y_c$. 
Only
for very high resolution, if $Q_c$ is lowered from 900 MeV
($y_c\sim 10^{-4}$) to the final $Q_0\sim 500$
MeV, about three quarters of the final multiplicity are
 produced, three times
as many jets as in the large complementary kinematic region with
$y_c> 10^{-4}$.
 The steep rise towards small $y_c$
 in Fig.~2 reflects the Landau singularity in the coupling
 at $y_\Lambda=\Lambda^2/Q^2$, which is
 however screened by the hadronization scale  $y_0=Q_0^2/Q^2\gsim
y_\Lambda$.
This behaviour is qualitatively described by the
high energy DLA result (\ref{ndla}) which shows the singularity at
$\lambda=0$. It will be interesting to study this behaviour experimentally,
but some additional assumptions on mass effects are necessary in this region
\cite{lo2}.

The prediction for the hadron multiplicity in gluon jets (Fig. 1b) is also
well met by experimental data at LEP energies \cite{lo2}; there are also
predictions for the dependence of the subjet multiplicity in gluon jets 
on the scale parameter $y_c$.

\end{document}